\documentclass[12pt]{iopart}

\usepackage{amssymb}
\usepackage{amstext}
\usepackage{amscd}
\usepackage{amsthm}
\usepackage{amsopn}
\usepackage{amsbsy}
\usepackage{eufrak}
\usepackage{indentfirst}
\usepackage{graphicx}

\newcommand{\be}{\begin{equation}}
\newcommand{\ee}{\end{equation}}
\newcommand{\bea}{\begin{eqnarray}}
\newcommand{\eea}{\end{eqnarray}}

\def\<{{\langle}}
\def\>{{\rangle}}

\begin{document}

 \title[Exponential sensitivity] {Exponential sensitivity of noise-driven switching in genetic networks}
\author{Pankaj Mehta }
\address{ Department  of Molecular Biology and Department of Physics, Princeton University, Princeton, NJ 08544}
\ead{pmehta@princeton.edu}
\author{Ranjan Mukhopadhyay}
\address{Department of Physics, Clark University,  Worcester, MA, 01610}

\author{Ned S.
 Wingreen} 
 \address{ Department of Molecular Biology, Princeton University, Princeton, NJ 08544}

\begin{abstract}
There is increasing experimental evidence that cells can utilize biochemical noise to switch probabilistically between distinct gene-expression states. In this paper,  we demonstrate that such noise-driven switching is dominated by tails of probability distributions and is therefore exponentially sensitive to changes in physiological parameters such as transcription and translation rates. Exponential sensitivity limits the robustness of noise-driven switching, suggesting cells may use other mechanisms in order to switch reliably. We discuss our results in the context of competence in the bacterium {\it Bacillus subtilis}.

\end{abstract}
\maketitle

\section{Introduction}

Recent experiments indicate that cells can use noise in gene expression to switch probabilistically between distinct gene-expression states \cite{Losick} .   For example, in the phenomenon of competence,  the soil-dwelling bacterium {\it Bacillus subtilis}   utilizes noise in gene expression to switch probabilistically between a vegetatively growing state and a competent state  where a bacterium can take-up exogenous DNA \cite{competence1a, competence1b, competence2a, competence2b, competence2c}. Noise is also likely responsible for the phenotypic heterogeneity found in mycobacteria \cite{Sureka} and the bistable behavior of {\it B. subtilis}  in nutrient rich media where genetically identical cells  are found in two distinct phenotypes: motile swimmers and immobile chains \cite{DubnauLosick}. Recent experiments also suggest that a noise-driven switch is at least in part responsible for the ability  of  both yeast \cite{Kaufmann} and the human fungal pathogen {\it Candida albicans} \cite{Zordan, Huang} to exist in two distinct phenotypes. The role of stochastic fluctuation in many other biological phenomenon such as the lytic-lysogeny decision during bactriophage infection is still the subject of current debate \cite{bactriophage1}-\cite{bactriophage3}.

In this paper,  we demonstrate that in standard models of noise-driven switching \cite{theory1} - \cite{Morelli} many biologically relevant quantities, such as the switching rate, exhibit an exponential sensitivity to physiological parameters.  The root cause of this exponential sensitivity is that switching properties are dominated by rare concatenations of  biochemical events \cite{ Roma, tenWolde}. Consequently, great care must be taken when calculating switching properties using coarse-grained models \cite{Morelli}.
In particular, we show that sources of noise often ignored in theoretical treatments, such a protein bursting \cite{burstingnoise1, burstingnoise2, burstingnoise3, burstingnoise4}, are  extremely important for noise-driven switches. However, provided mRNA lifetimes are short, we show that switching can still be accurately simulated using protein-only models of gene expression.

 The exponential sensitivity of switching may also have  important biological consequences. For example,  under many physiological conditions, {\it B. subtilis} cells can become competent only during a window of opportunity at the end of exponential growth \cite{competence1a, competence1b}.  Thus, in standard models of switching,  small changes in physiological parameter are likely lead to large variations in the percentage of cells that become competent.  It is then natural to ask if and how cells can achieve robust switching. We address this issue by discussing several alternative switching models that are more robust to changes in kinetics parameters.

 \section{Basic model of bistability}

\begin{figure}[t]
 \includegraphics[width=1.0\columnwidth]{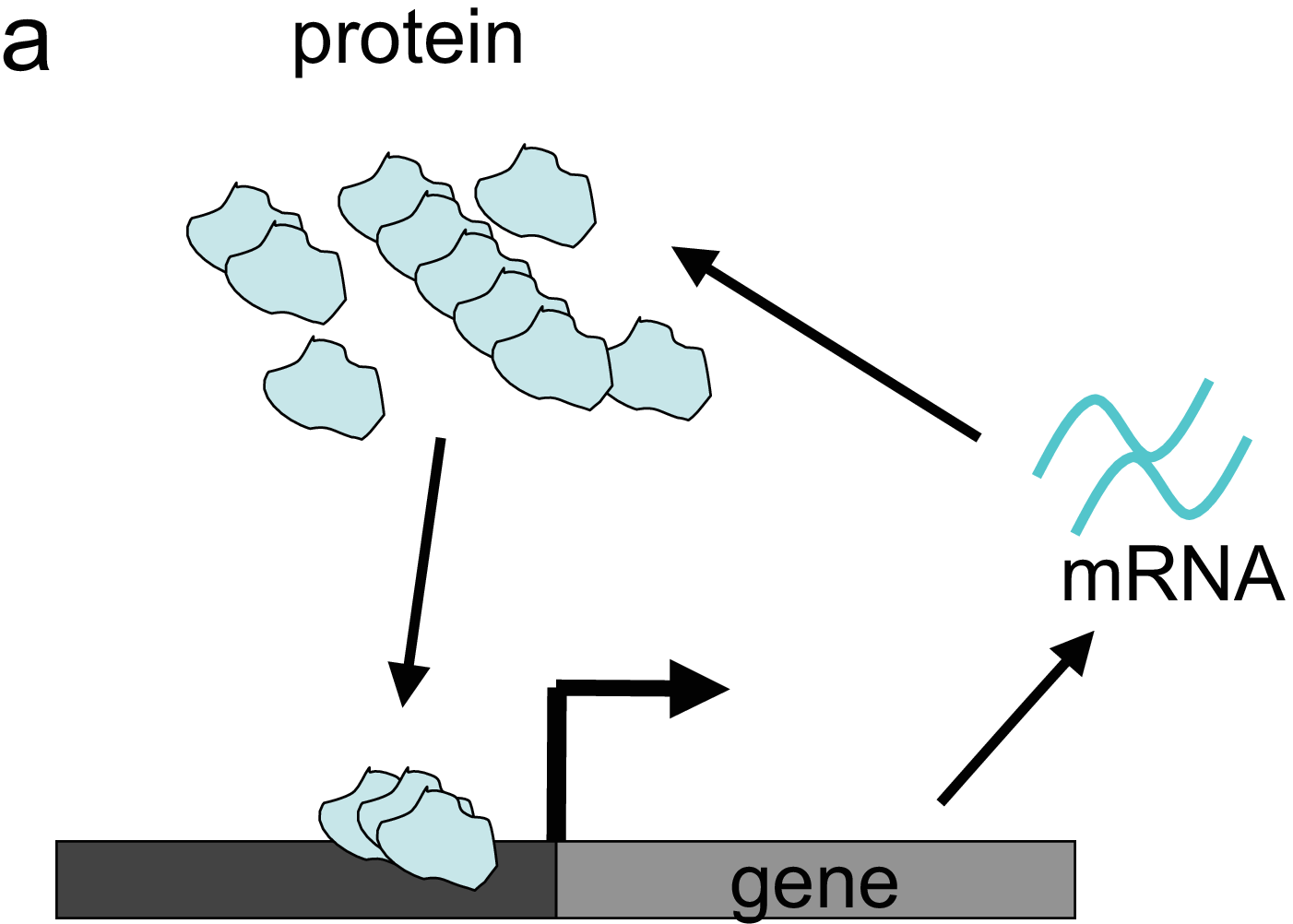}
\includegraphics[width=1.0\columnwidth]{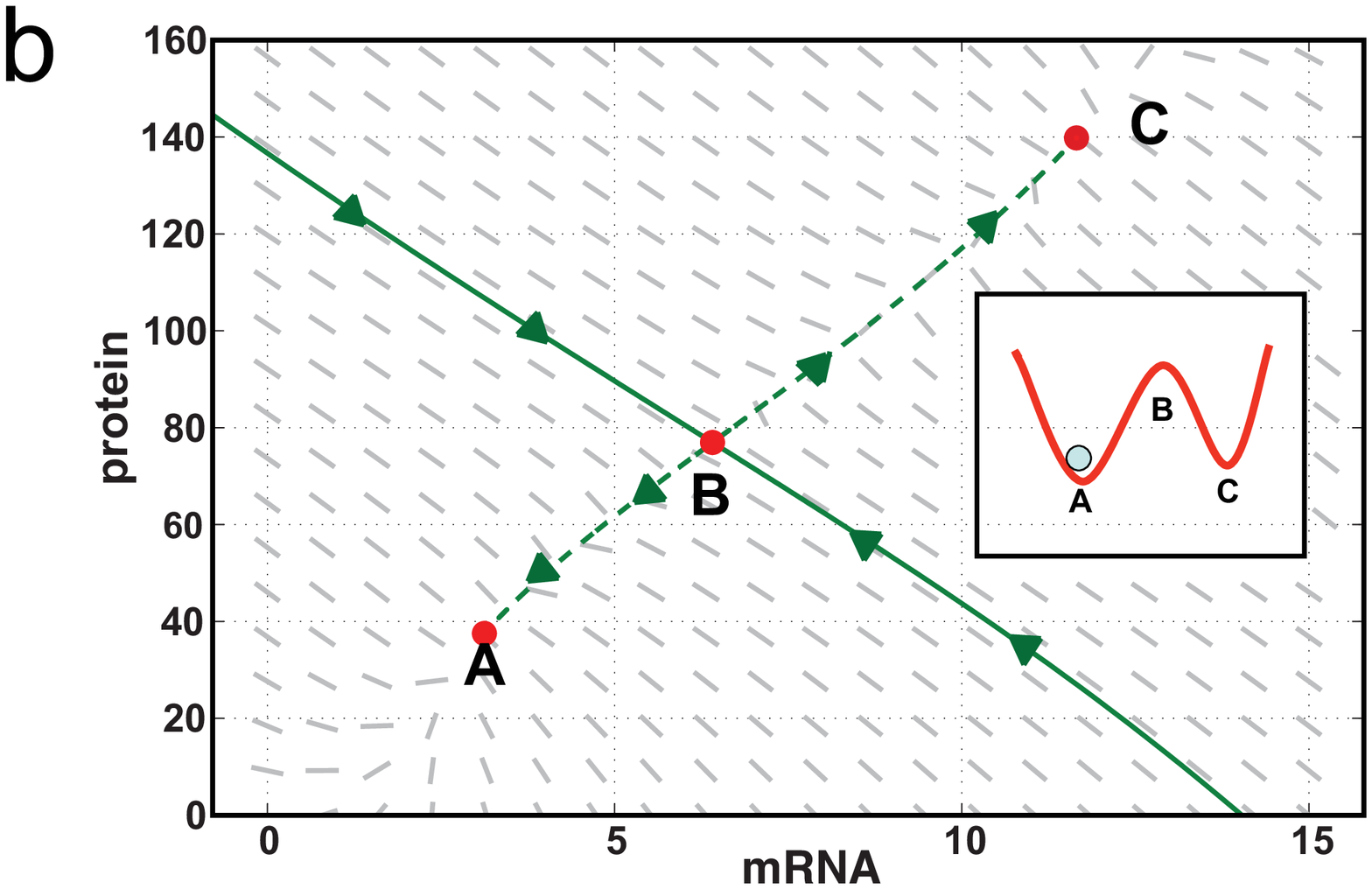}
\caption{(a) Schematic of a simple genetic network in which a protein activates its own transcription and (b) the phase portrait of the network dynamics. In mean-field theory,  this network can exhibit  bistable behavior with two stable fixed points (A and C), an intermediate unstable fixed point (B), and a separatrix (solid green curve) dividing the basins of attraction of the stable fixed points.  Noise in transcription (gene $\rightarrow$ mRNA) and translation (mRNA $\rightarrow$ protein) can drive transitions between A and C. Inset--
as discussed in the main text, switching is analogous to thermally-driven barrier-crossing in a double well potential.}
 \label{fig:figure1}
 \end{figure}

 Perhaps the simplest example of a genetic switch is a genetic network in which a protein induces its own expression (see Fig. \ref{fig:figure1}).  If this feedback is nonlinear, {\it  e.g.} if multiple copies of the protein are required to induce expression, such a network may be bistable with two distinct steady states, a low-expression state where the protein is present in small numbers and a high-expression state where the protein is present in large numbers.  Noise from the inherent stochasticity in biochemical reactions can switch probabilistically a cell between these two gene-expression states \cite{EMBOpaper1, EMBOpaper2}.  Such a simple bistable network is at the core of many noise-driven systems in biology \cite{DubnauLosick, Zordan, Huang}.   For example, in competence, the master-regulator protein ComK  promotes its own expression  through a positive-feedback loop with noise in gene expression switching bacteria from a low-ComK vegetative state to a  high-ComK competent state in response to stress. It is worth noting that under certain physiological conditions, the high-ComK competent state is actually transient. However, since we are concerned only with entrance into competence this will not be important  to our arguments below).

 The average number of mRNAs, $\bar{m}$,   proteins, $\bar{p}$,  in such a genetic network can be described by the deterministic differential equations
 \bea
\frac{d\bar{m}}{dt} &=& f_m(\bar{p})- \tau_m^{-1}\bar{ m}  
\label{meanequations1} \\
\frac{d\bar{p}}{dt}  &=& \alpha_p \bar{m} - \tau_p^{-1} \bar{p} 
\label{meanequations2}
\eea
where $f_m(p)$  is of the Hill form and is given by
\be
f_m(p) =  \alpha_{0m} + \frac{\alpha_m p^q}{K_d^q + p^q},
\ee
with $\alpha_{0m}$ the basal rate of  mRNA transcription and the second term representing positive feedback.
Eq. (\ref{meanequations1}) describes the change in the mean number of mRNAs due to mRNA transcription and degradation. Eq. (\ref{meanequations2}) describes the change in the mean number of proteins due to  protein translation and degradation. 
In the  bistable regime, these equations have two stable fixed points divided by a separatrix containing an unstable intermediate fixed point (see Fig. \ref{fig:figure1}). 

The deterministic differential equations (\ref{meanequations1}) and (\ref{meanequations2}) ignore fluctuations in mRNA and  protein  numbers. To capture the effects of these fluctuations, we employ the exact Master Equation for the probability, $P(m,p)$, of finding $m$ mRNA molecules and $p$ protein molecules \cite{VanKampen}. Define shift operators, $E_{s=m,p}^\pm$, which act on an arbitrary function $g(m,p)$ according to  $E_{m}^{\pm 1}g(m,p)=g(m \pm 1,p)$ and $E_p^{\pm 1} g(m,p)=g(m,p \pm 1)$. In terms of these operators, the Master Equation describing our simple bistable switch is 
\bea
\frac{\partial P(m,p ,t)}{\partial t}&=& [(E_p^{-1}-1)\alpha_p m + (E_p^{+1} -1)\tau_p^{-1}p \nonumber \\
&+& (E_m^{-1}-1)f_m(p)+ (E_m^{+1}-1)\tau_m^{-1}m]P(m,p,t).
\label{fullME}
\eea
The terms in Eq. (\ref{fullME}) have the same biological meaning as the corresponding terms in Eqs. (\ref{meanequations1}) and (\ref{meanequations2}). However, the Master Equation describes the full probability distributions $P(m,p)$ and not just the mean numbers of mRNAs and proteins. In writing this equation, we have assumed that the noise in the system is due to {\it intrinsic} noise in transcription and translation  \cite{burstingnoise1} and we have ignored other sources of noise.  In particular, we have modeled both protein and mRNA production with a simple Poisson model. This approximation may not hold in many biological systems where, in particular, transcription of mRNA may occur in bursts \cite{bindingnoise} -\cite{bindingnoise3}.

\section{Exponential Sensitivity of switching}

\begin{figure}[t]
\includegraphics[width=0.9\columnwidth]{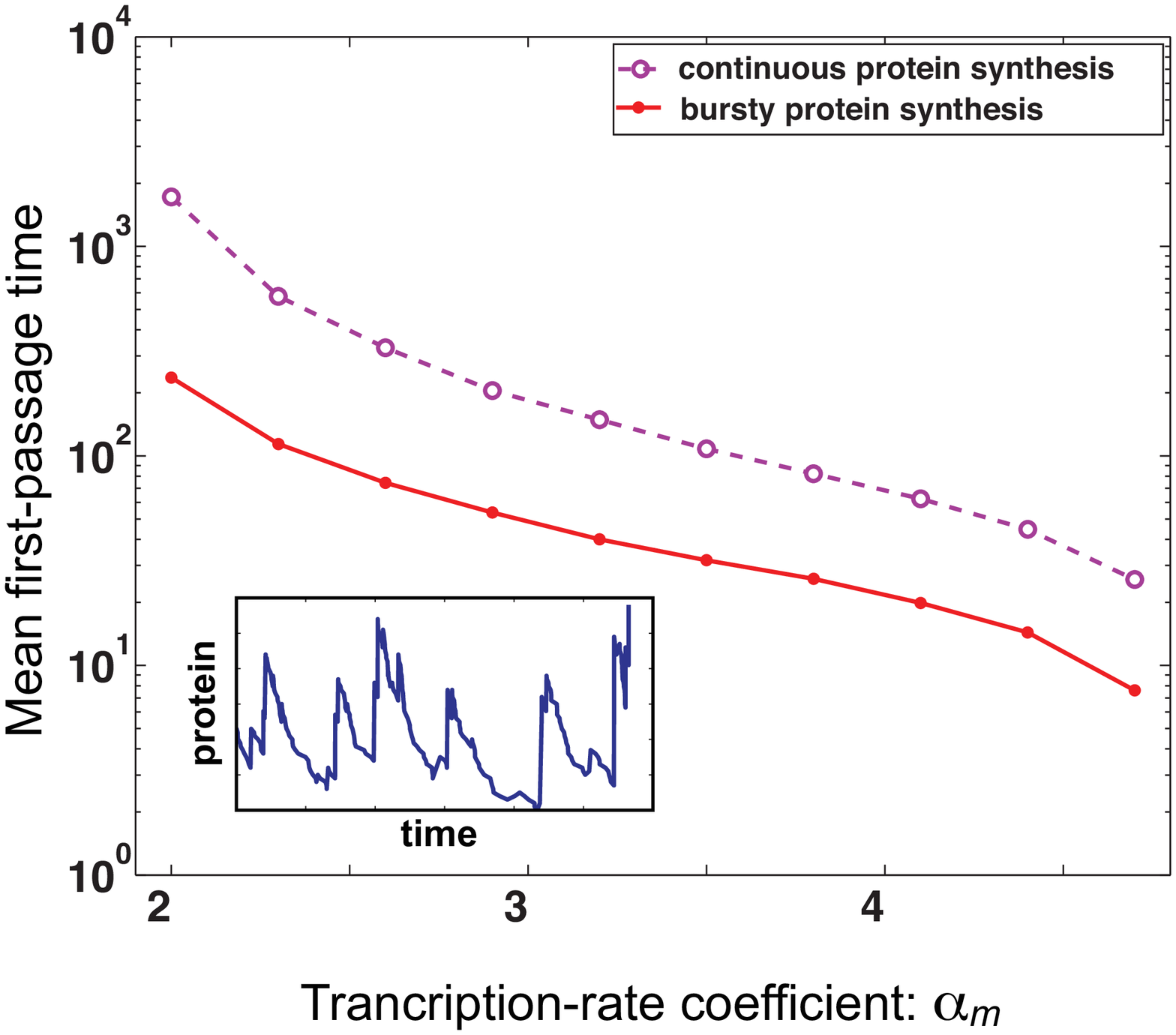}
\caption{ Mean first-passage time (MFPT) from the low-protein-number fixed point (A) to the separatrix (cf. Fig. 1) as a function of the transcription-rate coefficient  $\alpha_m$. The MFPT is shown for two distinct regimes: continuous protein synthesis  (dashed purple curve) where mRNA lifetimes and  protein lifetimes are equal  ($\tau_m/\tau_p=1$), and bursty protein synthesis(solid red curve)  where mRNA lifetimes are much shorter than protein lifetimes ($\tau_m/\tau_p=1/16$) so that proteins are made in bursts.   All times are in minutes and rates in inverse minutes. Each MFPT was obtained by averaging the first passage time from 5000 simulations, with parameters: $\tau_p=8$, $\alpha_{0m}=0.5$, $K_d=100$, $q=3$,  $\bar{b}=\alpha_p \tau_m= 10$.  Inset -typical stochastic trajectory for protein number from a Gillespie simulation with bursty protein synthesis with $\alpha_m=2.6$. }
\label{fig:figure2}
\end{figure}

An important property of biological switches is  the mean first passage time (MFPT) -- in our case,  the average time it takes to switch gene-expression states. The MFPT is also inversely related to the mean switching rate for a population. As discussed in the introduction, in competence the MFPT is directly related to the number of cells in a bacterial population that become competent during a window of opportunity at the end of exponential growth \cite{competence1a,  competence1b}. We have calculated the MFPT for the simple positive-feedback network described by  Eq. (\ref{fullME}) by performing 5000 independent Gillespie simulations  for each set of parameters, and averaging the first passage times (FPTs). (The Gillespie algorithm is a dynamic Monte-Carlo method that numerically generates trajectories of a stochastic Master Equation \cite{Gillespie}). The FPT for each simulation was calculated by initializing the system at the low-protein-number fixed point and calculating the time the trajectory took to reach the deterministic separatrix.
Figure \ref{fig:figure2} shows the MFPT for two different ratios of the mRNA and protein lifetimes, holding the mean protein number fixed.   When $\tau_m /\tau_p =1/16  \ll 1$, mRNAs are short-lived and typically many proteins are produced in a rapid burst from each mRNA molecule. In contrast, when $\tau_m/\tau_p=1$, proteins are produced nearly continuously. In the two simulations,  the mean number of proteins at the fixed points is identical. This is achieved by varying $\alpha_p$ with $\tau_m$ so that $\bar{b}=\alpha_p \tau_m$, the average number of proteins made per mRNA, is fixed.

 Notice that the MFPT is exponentially sensitive to the transcription-rate coefficient $\alpha_m$. Similar sensitivity is observed with respect to other parameters such as the protein-degradation rate (data not shown).  The MFPT in Fig. \ref{fig:figure2} also shows a strong dependence on the ratio of mRNA  and protein lifetimes though the mean protein number is held fixed. In particular,  the MFPT is significantly shorter when proteins are produced in bursts ($\tau_m/\tau_p \ll 1$). Bursty protein production increases protein fluctuations, and as a result, lowers the MFPT for switching. This dramatic reduction in MFPT due to bursting has also been observed in the context of a dimerizing genetic switch \cite{Warren}.The increase in protein fluctuations due to bursts of production can be quite large. For example, the variance of the number of proteins for a fixed rate of mRNA transcription is given in the bursting limit by \cite{burstingnoise1, burstingnoise2, burstingnoise3, burstingnoise4} 
\be
\< \delta p^2 \>/\bar{p}^2= \frac{1 +\bar{b} }{\bar{p}},
\label{bnoise}
\ee
where $\bar{b}=\alpha_p \tau_m \gg 1$ is  the mean burst size, {\it i.e.} the mean number of proteins made from each mRNA. This should be contrasted with the case where proteins are produced continuously and $ \< \delta p^2 \>/\bar{p}^2=1/\bar{p} \ll (1+\bar{b})/\bar{p}$.

  In the bursting limit ($\tau_m \ll \tau_p$), gene regulation is often approximated using protein-only models. In these approximations, time-derivatives of the mRNA species are set equal to zero, and the resulting mean  mRNA concentrations are substituted into the equations describing protein dynamics.  
For our network, such a procedure yields the reduced equation
\be
\frac{d\bar{p}}{dt} =  \alpha_p \bar{m}-\tau_p^{-1}\bar{p} = \bar{b}f_m(\bar{p})-\tau_p^{-1} \bar{p}.
\label{onespecmeanequation}
\ee
It is worth noting that  since these approximations  fail when $\tau_m \ge \tau_p$,  switching properties cannot  be accurately described by protein-only models in the continuous protein synthesis regime. 

\section{Incorporating translational bursting in protein-only models}

\begin{figure}[t]
\includegraphics[width=0.9\columnwidth]{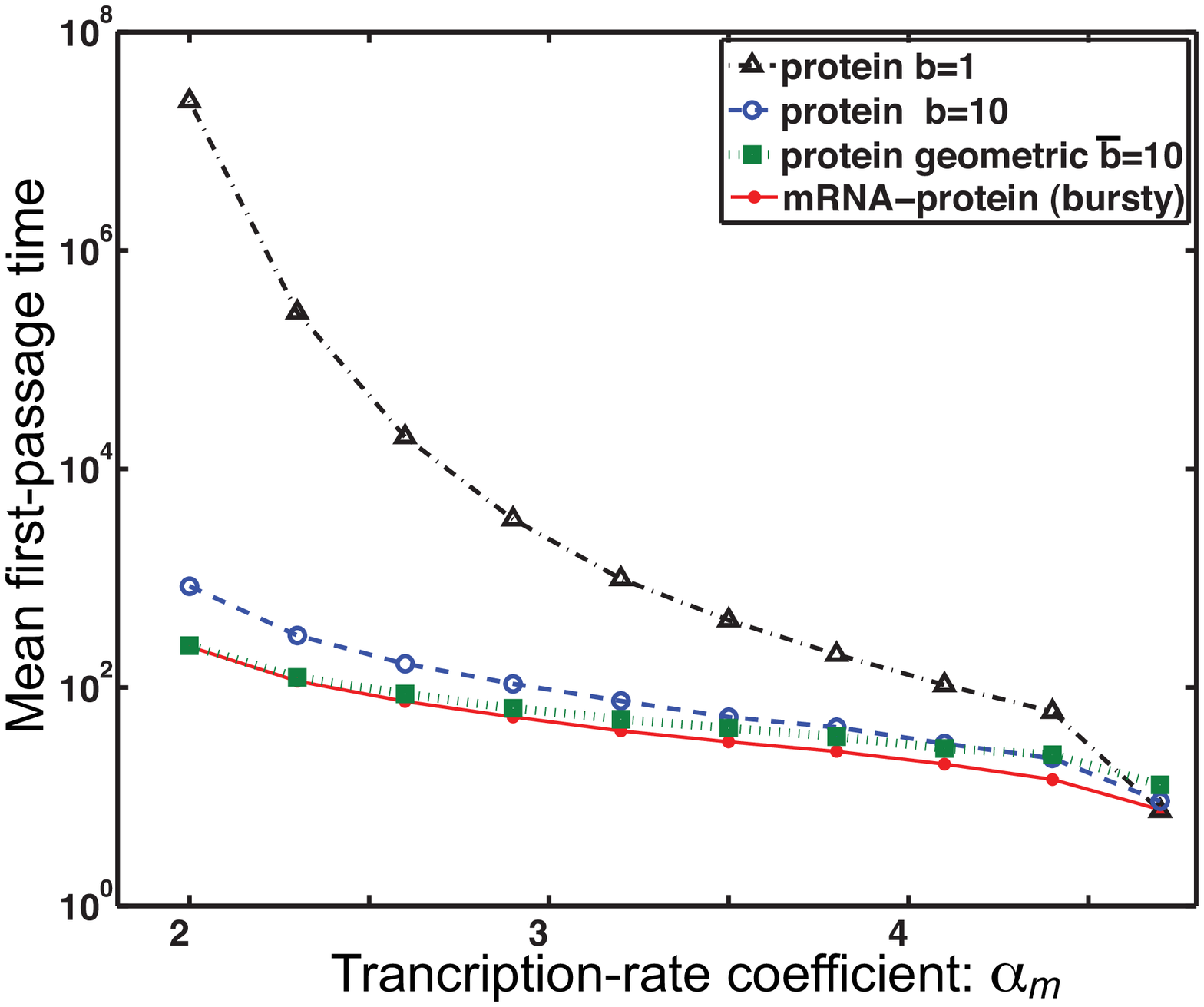}
\caption{Comparison of mean first-passage times (MFPTs) between the full mRNA-protein model and various protein-only models. The mean protein level at the fixed point is the same for all models and the MFPT is the average of $5000$ independent simulations: (solid red curve) MFPT for bursty mRNA-protein model from Fig. 2; (dotted green curve) MFPT for a model where proteins are produced in bursts of size $b$, where $b$ is drawn from a geometric distribution with
mean $\bar{b}=10$; (dashed blue curve) MFPT for a protein-only model where proteins are produced in bursts of size $b=10$; (dot-dashed curve) MFPT for a model where proteins are produced without bursting ($b=1$). }
\label{fig:figure3}
\end{figure}

Even in the bursting regime, commonly used protein-only models often fail to accurately reproduce switching kinetics since these models neglect or oversimplify noise sources such as protein bursting \cite{Warren}.
Stochastic fluctuations are often included in protein-only models by assuming that proteins
are produced probablistically one at a time. This assumption leads to the Master Equation
\be
\frac{\partial P(p,t)}{\partial t} = [(E_p^{-1}-1)\bar{b}f_m(p)+(E_p^{+1}-1)\tau_p^{-1} p]P(p,t).
\label{onespecME}
\ee     
We have calculated the MFPT for switching between the low-expression and high-expression states
for this Master Equation using the Gillespie algorithm and the result is shown in  Fig. \ref{fig:figure3}.  Notice that the MFPTs calculated in this way can be nearly four orders of magnitude longer than those calculated using the full mRNA-protein  model, demonstrating that protein bursting significantly increases the rate of switching.

In fact it is straightforward to include protein bursting in protein-only models in the bursting regime \cite{Friedman}.  Since mRNA degradation and protein translation from an mRNA molecule are independent Poisson processes, it can be easily shown that the protein-burst size $b$ (defined as the number of proteins produced from a single mRNA molecule) is geometrically distributed with mean burst size $\bar{b}=\alpha_p \tau_m$ and distribution
\be
G(b)= \frac{1}{1+\bar{b}}\left(\frac{\bar{b}}{1+\bar{b}}\right)^b.
\ee 

We consider two different approximations for incorporating bursting in a protein-only model. In the first, once an mRNA is produced, translation proceeds deterministically and $\bar{b}$ proteins are produced. In the second, $b$ proteins are produced  from a single mRNA molecule  with the geometric distribution, $G(b)$.  Both these  approximations assume $\tau_m \ll \tau_p$ since all the proteins from a single mRNA molecule are considered to be produced instantaneously. The MFPTs  computed within these two  approximations are shown in  Fig. \ref{fig:figure3}. As for the full mRNA-protein model, the protein-only MFPTs are exponentially sensitive to changes in the transcription-rate coefficient $\alpha_m$.  The  MFPT computed using the geometric distribution of protein-burst sizes agrees well with the MFPT from the full  mRNA-protein model. However, the protein-only  model where proteins are produced with a fixed burst size $\bar{b}$ is inadequate for calculating the MFPT even though this approximation correctly yields Eq. (\ref{bnoise}) for the variance in protein number at a fixed rate of mRNA transcription. Note that the MFPT computed within the geometric-burst-size approximation is smaller than that computed using the fixed-burst-size approximation. This is expected since geometrically distributed burst sizes increase the fluctuations in protein production.

\section{Discussion}

 A qualitative understanding of  our simulation results  can be obtained from the Fokker-Planck approximation to the  Master Equation (\ref{onespecME}). Gaussian approximations such as the Fokker-Plank and Langevin equations generally fail to accurately capture switching properties since they underestimate rare fluctuations \cite{Roma}, but nonetheless they allow us to identify the regime of exponential sensitivity. The Fokker-Plank equation that approximates our genetic network
is  \cite{VanKampen}: 
  \be
\frac{\partial P(p,t)}{\partial t}= -\frac{\partial}{\partial p} 
[a_1(p)P(p,t)] + \frac{1}{2} \frac{\partial^2}{\partial p^2}[a_2(p)P(p,t)],
\label{FP}
\ee
where $a_1(p) = \bar{b}f_m(p)-\tau_p^{-1} p$ and $a_2(p) = \bar{b}^2 f_m(p) + \tau_p^{-1} p$.
In deriving (\ref{FP}),  we assumed that the state of the system is described by a single reaction coordinate, protein number, and have ignored fluctuations such as transcription-factor binding and unbinding. Additionally, we have made the simplifying approximation that proteins are produced
in bursts of size $\bar{b}$.We can calculate the MFPT using Eq. (\ref{FP}) and obtain  \cite{ FPEMFPT}: 
\be
{ \rm MFPT} \propto \exp{ \left(\frac{U(\bar{p}_B) -U(\bar{p}_A)}{a_2(\bar{p}_A)}\right) }\equiv \exp{\left(\frac{U_{BA}}{D}\right)}.
\ee 
where we have defined an effective potential $U(p)$ using the equation $\frac{\partial U}{\partial p}=-a_1(p)$ and where
$\bar{p}_{A(B)}$  is the mean number of proteins at the low-expression stable fixed point (intermediate-expression, unstable fixed point). The result is derived assuming the constant-diffusion approximation where we replace $a_2(p)$ by its value at the low-expression fixed point $a_2(\bar{p}_A)$ and performing a saddle-point approximation. As expected,  this approximate Fokker-Plank equation correctly determines the regime of exponential sensitivity but does not quantitatively reproduce the MFPT. Notice that the MFPT resembles the FPT for  barrier crossing by a particle moving in a potential $U(p)$ with effective diffusion constant $D=a_2(\bar{p}_A)$ and barrier height $U_{BA}$ \cite{Kramer}.  
The analogy between barrier crossings and biochemical switches was discussed earlier in  \cite{theory6, theory6b}. We have extended this analogy to include protein bursting and shown that the burst size changes both the barrier height and the effective diffusion constant.  When $U_{BA}/D \gg 1$,{\it i.e.} when the  barrier height is large,  the MFPT is exponentially sensitive to changes in the kinetic parameters that determine $U(p)$ and $D$.  Thus, exponential sensitivity indicates that switching is dominated by the tails of a probability distribution and the central limit theorem no longer applies.  Thus,  aspects  of gene regulation such as protein bursting which may not be important when considering small fluctuations around mean protein levels, may still play an important role in governing switching rates.

\section{Conclusion and Outlook}

 There is increasing evidence that the dynamic properties of bistable genetic switches are extremely sensitive to the details of gene regulation. This work, as well as others \cite{Morelli0, Morelli, Warren}, show that small changes in parameters can lead to large changes in the dynamics of switching. We have argued that this sensitivity occurs because the dynamics of switching depends exponentially on the effective barrier height and the effective diffusion coefficient for switching. Such exponential sensitivity has important theoretical implications for coarse grained descriptions of biochemical networks \cite{Morelli0}. For example, many sources of noise such as mRNA bursting \cite{bindingnoise} -\cite{bindingnoise3} and protein dimerization \cite{theory2, Morelli, Morelli0},  likely play important roles in determining switching rates in some biological systems. Thus, great care must be taken theoretically when integrating out sources of noise \cite{Morelli0}. This work complements and extends earlier 
theoretical studies by showing that when mRNA lifetimes are short, protein fluctuations can be correctly incorporated using protein-only models. In addition, it provides a simple understanding of exponential sensitivity using intuition from the statistical mechanics of barrier crossings.

The exponential sensitivity of genetic switches also has important physiological consequences.
In recent experiments,  Maamar {\it et al.}  \cite{competence1a,  competence1b} showed that protein bursting is a major source of the noise  driving the induction of competence. Furthermore, they were able to change the noise-driven switching rate (the inverse of the MFPT)  in the {\it B. subtilis} competence system, while keeping mean protein numbers fixed, by   manipulating transcription and translation rates.  This suggests that evolution can  tune switching rates while keeping mean protein levels fixed.   (Note, however that switching rates are also likely to be influenced by additional factors such as  temperature, pH, fluctuations in the number of cellular components important for transcription and translation such as polymerases and ribosomes, and extra-cellular factors such as chemicals present in the environment.)

By our analysis, switching -- because it is like barrier crossing in statistical mechanics -- is likely to be exponentially sensitive to all these factors. Indeed, this may be one of the causes for the large variations observed in experiments on the number of swimming cells and chains in a cultures of {\it B. subtilis} \cite{DubnauLosick}. However, it is then natural to ask if, and how, cells can achieve robust switching. For example, robust switching may be achieved using alternative switching mechanisms such as  all-or-none events that switch the cell between the low- and high-expression states \cite{Novick}. Such an all-or-none event could correspond to the disassembly of an entire repressor complex  bound to the promoter of a gene, allowing multiple mRNA molecules to be made rapidly. If switching is dominated by these all-or-none events and not the tails of probability distributions, then switching is no longer  necessarily exponentially sensitive to changes in parameters. This scenario may be relevant to competence since a repressor-complex is known to be important in regulating the master competence regulator ComK \cite{Smits2}. Alternatively, cells could switch robustly by actively tuning noise to control switching rates. Recent experiments suggest that this may indeed be the case in mycobacteria \cite{Sureka}. A final possibility is that large variations in the switching rate may be acceptable or even desirable for cells living in certain environments \cite{Kussell1, Kussell2}.  Thus, cells may exploit the exponential sensitivity of noise-driven switching to increase their fitness.  It will be interesting to better understand the ecological and evolutionary implications of noise-driven switching.

\section{Acknowledgments}

 We would like to acknowledge D. Dubnau, H. Maamar, A. Raj, M. Elowitz, G. Suel, S. Goyal, and A. Sengupta for useful discussions. This work was supported in part by  US National Institutes of Health (NIH) grants R01 GM073186 and PSO  GM071508, and by the Defense Advanced Research Project Agency (DARPA) under grant HR0011-05-0057.

\section{Glossary}

Competence - A physiological state that allows a bacterial cell to take up exogenous DNA. \\
\\
Master Equation - A mathematical equation that describes the time evolution of a probability distribution. \\
\\ 
Mean First Passage Time (MFPT) - In our context, the average time it takes for a bistable genetic switch to transition from one gene-expression state to another due to fluctuations. The MFPT is inversely related to the switching rate. \\
\\

\end{document}